\documentclass{article}
\usepackage{amsmath}

\newcommand{\vekt}[1]{\mbox{\boldmath $#1$\unboldmath}}
\renewcommand{\theequation}{\thesection.\arabic{equation}}
\input{tcilatex}

\begin{document}

\title{Mean Velocity Equation for Turbulent Fluid Flow: \\
An Approach via Classical Statistical Mechanics}
\author{J\"{u}rgen Piest \\
Meisenweg 13\\
D-24147 Klausdorf, Germany\\
piestj@aol.com\\
0049-431-791297}
\maketitle

\begin{abstract}
The possibility to derive an equation for the mean velocity field in
turbulent flow by using classical statistical mechanics is investigated. An
application of projection operator technique available in the literature is
used for this purpose. It is argued that the hydrodynamic velocity defined
there, in situations where the fluid is turbulent, is to be interpreted as
the mean velocity field; in that case, the momentum component of the
generalized transport equation derived there is the mean velocity equation.

In this paper, stationary incompressible flow for constant mass density and
temperature is considered. The stress tensor is obtained as a nonlinear
functional of the mean velocity field, the linear part of which is the
Stokes tensor. The formula contains a time correlation function in local
equilibrium. Presently, there exists a microscopic theory for time
correlations in total equilibrium only. For this reason and as a preliminary
measure, the formula has been expanded into a power series in the mean
velocity; though this limits the applicability to low Reynolds number flow.
The second order term has been evaluated in a former paper of the author.
For the third order term, the form of the kernel function is derived. Its
calculation with the aid of the mode-coupling theory is completed; it will
be reported in an separate paper. An numerical application with the data of
the circular jet is under way.

\end{abstract}

\section{Introduction}

In turbulence experiments, very often the relevant variables - those which
can be related to the external conditions of the experiment - are not the
actual quantities but their mean values. The process is then described by a
statistical theory. It is well known that it can be very difficult to derive
equations for mean values which are closed. It can be argued that the reason
is that it is difficult to construct the multi-point probability
distribution for the process.

In this paper, the possibility to derive an equation for the mean velocity
field in turbulent flow by using classical statistical mechanics is
investigated. Then, in principle the probability distribution of the process
can always be constructed starting from the total equilibrium distribution.
On the other hand, it will be necessary to distinguish between \ \ \ \ \ \ \
\ macroscopic and microscopic parts of the motion and to formulate the
latter in a way suitable for hydrodynamic purpose.

It is known that the Navier-Stokes equation can be derived from Statistical
Mechanics. This has been performed first by Chapman and Enskog in 1916 and
1917, with the Boltzmann equation as a starting point; see, e. g., Huang %
\cite{hu}. More recently, a formalism has been developed which starts from
the Liouville equation and applies projection operator technique; see
Zwanzig \cite{zw60}, Mori \cite{mo65}. Here, the presentation of \ Grabert %
\cite{gr} is used as a reference. In the relevant part of this work, a
generalized transport equation is derived; the application for simple fluids
and a suitable approximation of the stress tensor lead to the Navier-Stokes
equation. In the present paper, arguments are given that for situations
where the fluid flow is turbulent, the momentum component of the generalized
transport equation for simple fluids actually is the mean velocity equation.
The formulation is restricted to stationary flow in incompressible fluid of
constant density and temperature.

In section 2, some definitions of Statistical Mechanics and the basic steps
which lead to the generalized transport equation are referenced from \cite%
{gr}. The interpretation of the momentum component as the mean velocity
equation is explained. It is known that the form of this equation is very
similar to the Navier-Stokes equation, with an additional friction force
term (Reynolds force). There are numerous approaches to formulate this
quantity (see, e. g., \cite{mc} and the references therein); it is seen that
it is nonlinear in the velocity and, after all evidence, also nonlocal. - As
a preliminary measure, the friction force has been expanded into a power
series in the velocity; though this limits the application to low Reynolds
number flow. This is explained in section 3. The main results for the $%
2^{nd} $ order term reported earlier \cite{pi89}, \cite{pi90} \ are quoted;
the calculation of the formula for the $3^{rd}$ order term is presented in
this paper.

\section{Mean velocity equation}

In this section, the definitions and the basic steps of the derivation of
the generalized transport equation are referenced from Grabert \cite{gr};
the notation is slightly different. The fluid is considered to be a system
of $N$ particles of mass $m$ with positions $%
\vekt{y}%
_{j}$ and velocities $%
\vekt{v}%
_{j}$ (simple fluid) which are combined to the phase space vector $%
\vekt{z}%
$. Vector components are described by Latin indices, e. g. $%
\vekt{y}%
_{j}=\{y_{ja}\}$ . The particles interact with a symmetric, short-ranged,
pairwise additive interparticle potential. The system is enclosed in a box
of Volume $V$ . A function $g(%
\vekt{z}%
)$ is called a phase space function, or microscopic variable. Especially, we
need the space densities of the conserved quantities particle number, energy
and momentum $n$, $e$, $\mathbf{%
\vekt{p}%
}$ which are collected to a 5-element linear matrix $%
\vekt{a}%
$ . They are functions of an additional space variable $%
\vekt{x}%
$:%
\begin{equation}
\vekt{a}%
=\sum_{j=1}^{N}\widetilde{%
\vekt{a}%
}_{j}\delta (%
\vekt{x}%
-%
\vekt{y}%
_{j})  \label{2.1}
\end{equation}

For the particle functions $\widetilde{%
\vekt{a}%
}_{j}$ we have $\widetilde{n}_{j}=1$ , $\widetilde{\mathbf{%
\vekt{p}%
}}_{j}=m%
\vekt{v}%
_{j}$ , while the energy function contains the interparticle potential. The
quantities $%
\vekt{a}%
$ obey the conservation relations:%
\begin{equation}
\overset{\cdot }{%
\vekt{a}%
}=-\nabla \cdot 
\vekt{s}
\label{2.2}
\end{equation}

The fluxes $%
\vekt{s}%
$ have the same general structure as the $%
\vekt{a}%
$ (\ref{2.1}); the particle functions can be found in \cite{gr}; especially,
we have $%
\vekt{s}%
_{1}=\mathbf{%
\vekt{p}%
/}m$ . The time evolution of any phase space function $g$ is described by
the Liouville equation: 
\begin{equation}
\overset{\cdot }{g}\,=i\mathcal{L}g  \label{2.2.1}
\end{equation}

$i\mathcal{L}$ is the Liouville operator, a linear differential operator the
form of which can be found in any textbook of statistical mechanics. From (%
\ref{2.2.1}), the formal solution for $g(t)$ given the initial value $g$ is:%
\begin{equation}
g(t)=\func{e}^{i\mathcal{L}t}g  \label{2.2.2}
\end{equation}

In the statistical model, $%
\vekt{z}%
$ and $N$ are considered random variables; that is, the \ probability
density $f(%
\vekt{z}%
,N)$ is of grand canonical type. The ensemble mean value (expectation) of a
phase space function $g$ is defined in the `Heisenberg' picture:%
\begin{equation}
\langle g\rangle (t)=\sum_{N=1}^{\infty }\int d%
\vekt{z}%
g(%
\vekt{z}%
,N,t)f(%
\vekt{z}%
,N)  \label{2.3}
\end{equation}

In this formula, $f(%
\vekt{z}%
,N)$ is the initial probability distribution, and $g(%
\vekt{z}%
,N,t)$ is the value of $g$ at time $t$ if the initial positions and
velocities of the particles are described by $%
\vekt{z}%
$ . The operation (integration + Summation) is sometimes indicated by the
symbol `tr':%
\begin{equation}
\func{tr}\{\Omega \}=\sum_{N=1}^{\infty }\int d%
\vekt{z}%
\Omega (%
\vekt{z}%
,N)  \label{2.4}
\end{equation}

Certain probability densities (also called distributions here) are
frequently used in the analysis. One of them is the (total) equilibrium
distribution which corresponds to macroscopic rest: 
\begin{subequations}
\label{2.5}
\begin{gather}
f_{0}=\psi (N)\exp (\Phi _{0}+\beta (\mu N-H(%
\vekt{z}%
)))  \label{2.5a} \\
\psi (N)=\frac{1}{N!}(\frac{m}{h})^{3N}  \label{2.5b}
\end{gather}

Here, $h$ is Planck's constant, $\beta =1/(k_{B}T)$ , $k_{B}$ being
Boltzmann's constant and $T$ the temperature, $\mu $ is the chemical
potential which is a function of mass density and temperature, and $H(%
\vekt{z}%
)$ is Hamilton's function which describes the total energy of the fluid. For
the normalization constant, we have$\ \Phi _{0}=-\beta PV$, $P$ being the
equilibrium pressure. Expectations with respect to the equilibrium
distribution are denoted by $\langle \rangle _{0}$. In case of a simple
fluid, the 'relevant probability distribution' of Grabert's formalism (see %
\cite{gr}, sec. 2.2) is the local equilibrium distribution: 
\end{subequations}
\begin{subequations}
\label{2.6}
\begin{gather}
f_{L}(t)=\psi (N)\exp (\Phi (t)-%
\vekt{a}%
(%
\vekt{z}%
)\ast 
\vekt{b}%
(t)),  \label{2.6a} \\
\vekt{b}%
=\{\beta (\frac{m}{2}u^{2}-\mu ),\beta ,-\beta \mathbf{%
\vekt{u}%
\},}  \label{2.6b} \\
\Phi (t)=-\log (\func{tr}\{\psi \exp (-%
\vekt{a}%
\ast 
\vekt{b}%
(t))\}).  \label{2.6c}
\end{gather}

Here the symbol $\ast $ is introduced for the operation: Multiplication,
plus Summation over the 5 elements of the linear matrices $%
\vekt{a}%
$ , $%
\vekt{b}%
$, plus Integration over geometrical space. The elements of $%
\vekt{b}%
$ are called the conjugate parameters; they are functions of the quantities $%
\beta $ , $\mu $ and $\mathbf{%
\vekt{u}%
}$ which we will sometimes call the thermodynamic parameters. $\beta =1/(kT)$%
, with $k$ being Boltzmann's constant and $T$ the absolute temperature; $\mu 
$ is the chemical potential which is a function of temperature and pressure,
and $\mathbf{%
\vekt{u}%
}$\ is the hydrodynamic velocity. These quantities will be considered slowly
varying functions of space and time. The $%
\vekt{b}%
$\ are defined such that the expectations of the $%
\vekt{a}%
$ are identical to their expectations in local equilibrium: 
\end{subequations}
\begin{equation}
\langle 
\vekt{a}%
\rangle =\langle 
\vekt{a}%
\rangle _{L}  \label{2.7}
\end{equation}

The projection operator techniqe (POT) is a means for separating macroscopic
and microscopic parts of the random variables. It starts by defining the set
of phase space functions which are relevant for the description of the
process. For simple fluids, this set is identified with the densities of
conserved variables, $%
\vekt{a}%
$\ . A projection operator \ is defined which projects out of any
microscopic variable $g$ the part which is proportional to the relevant
variables. It reads:

\begin{equation}
\mathcal{P}g=\langle g\rangle _{L}+\langle g\,\delta 
\vekt{a}%
\rangle _{L}\ast \langle \delta 
\vekt{a}%
\,\delta 
\vekt{a}%
\rangle _{L}^{-1}\ast \delta 
\vekt{a}
\label{2.8}
\end{equation}

Here, $\delta 
\vekt{a}%
=%
\vekt{a}%
-\langle 
\vekt{a}%
\rangle _{L}$; $\langle \rangle _{L}^{-1}$denotes the inverse of the
expectation matrix in the formula. For stationary flow, $\mathcal{P}$ is
time independent. The analysis in \cite{gr} starts by splitting the
exponential operator (\ref{2.2.2}):%
\begin{equation}
\func{e}^{i\mathcal{L}t}=\func{e}^{i\mathcal{L}t}\mathcal{P+}%
\int_{0}^{t}dt^{\prime }\func{e}^{i\mathcal{L}t^{\prime }}\mathcal{P}i%
\mathcal{L}(1-\mathcal{P})\func{e}^{(1-\mathcal{P})i\mathcal{L}(t-t^{\prime
})}+(1-\mathcal{P})\func{e}^{(1-\mathcal{P})i\mathcal{L}t}  \label{2.9}
\end{equation}

This corresponds to \cite{gr}\ , formula (2.4.1), specialized to stationary
flow, where especially $\mathcal{P}=$const$(t)$. With (\ref{2.9}), the
Liouville equation (\ref{2.2.1}) is reformulated:%
\begin{multline}
\overset{\cdot }{%
\vekt{a}%
}=\func{e}^{i\mathcal{L}t}\mathcal{P}i\mathcal{L}%
\vekt{a}%
\mathcal{+} \\
+\int_{0}^{t}\,\,dt^{\prime }\func{e}^{i\mathcal{L}t^{\prime }}\mathcal{P}i%
\mathcal{L}(1-\mathcal{P})\func{e}^{(1-\mathcal{P})i\mathcal{L}(t-t^{\prime
})}i\mathcal{L}%
\vekt{a}%
+(1-\mathcal{P})\func{e}^{(1-\mathcal{P})i\mathcal{L}t}i\mathcal{L}%
\vekt{a}
\label{2.10}
\end{multline}

By averaging over the initial probability density, and after some
manipulations, Grabert's generalized transport equation \ \cite{gr},
(2.5.17) ist obtained. Below, this equation is presented for stationary flow
in simple fluids. It is postulated in POT that the initial probability
density is of the form of the relevant probability density. Grabert states
that this should not be considered a general restriction of the method but a
means to form the general particle system into the type specially
considered; see \cite{gr}, sec. 2.2 . In \cite{gr}, sec. 8.3, it is shown
that for simple fluids the relevant probability density is that of local
equilibrium. For the present approach this means that turbulent flow is
considered which initially developed from laminar flow with suitable
velocity gradient. - It is a consequence of this postulate that the last
term in (\ref{2.10}) vanishes after averaging. Moreover, it is shown that:%
\begin{equation}
\left\langle \func{e}^{i\mathcal{L}t}\mathcal{P}i\mathcal{L}%
\vekt{a}%
\right\rangle =\left\langle \overset{\cdot }{%
\vekt{a}%
}\right\rangle _{L}=-\nabla \cdot \left\langle 
\vekt{s}%
\right\rangle _{L}  \label{2.11}
\end{equation}

In the last step, the conservation relations (\ref{2.2}) are introduced.
Stationary flow is considered to be the process described by the generalized
transport equation under stationary conditions and for very large times. One
obtains:%
\begin{equation}
0=-\nabla \cdot \left\langle 
\vekt{s}%
\right\rangle _{L}+%
\vekt{D}
\label{2.12.1}
\end{equation}
\begin{equation}
\vekt{D}%
_{\alpha }(%
\vekt{x}%
)=-\nabla _{c}\int d%
\vekt{x}%
^{\prime }\int_{0}^{\infty }dt\left\langle [\func{e}^{(1-\mathcal{P})i%
\mathcal{L}t}\widehat{s}_{\alpha c}(%
\vekt{x}%
)]\widehat{s}_{\beta d}(%
\vekt{x}%
^{\prime })\right\rangle \nabla _{d}^{\prime }b_{\beta }(%
\vekt{x}%
^{\prime })  \label{2.12.2}
\end{equation}%
\begin{equation}
\widehat{s}_{\alpha c}(%
\vekt{x}%
)=(1-\mathcal{P})s_{\alpha c}(%
\vekt{x}%
)  \label{2.12.3}
\end{equation}

Latin and greek indices run over 3 and 5 values, respectively; this is
sometimes expressed by saying that, e. g., the index $\alpha $\ 'runs over
1, 2 and the latin index $a$'. Eqs. (\ref{2.12.1}), (\ref{2.12.2})
correspond to \cite{gr}, (8.1.13), (8.5.1) for stationary flow. From these,
now the momentum equation is taken, for which one obtains (\cite{gr},
(8.4.15)): 
\begin{equation}
\left\langle 
\vekt{s}%
_{ac}\right\rangle _{L}=\rho 
\vekt{u}%
_{a}%
\vekt{u}%
_{c}+P\delta _{ac}  \label{2.13}
\end{equation}

Here, $\delta _{ac}$\ is the Kronecker symbol, $\rho $\ the mass density and 
$%
\vekt{u}%
$ the fluid velocity defined by:\ 
\begin{equation}
\left\langle 
\vekt{p}%
\right\rangle =\rho 
\vekt{u}
\label{2.14}
\end{equation}

(\cite{gr}, (8.3.12)). At this point, the continuity equation is introduced.
This actually is the mass density component of \ (\ref{2.12.1}), and for
incompressible constant density flow it reduces to:%
\begin{equation}
\nabla \cdot 
\vekt{u}%
=0  \label{2.15}
\end{equation}

With these formulas and (\ref{2.6b}), (\ref{2.12.1}), (\ref{2.12.2}) obtain
their final form:

\begin{equation}
\rho \mathbf{%
\vekt{u}%
\cdot \nabla 
\vekt{u}%
=-\nabla }P+\mathbf{\nabla }\cdot 
\vekt{R}
\label{2.16}
\end{equation}%
\begin{equation}
R_{ac}(%
\vekt{x}%
)=\int d%
\vekt{x}%
^{\prime }S_{abcd}(%
\vekt{x}%
,%
\vekt{x}%
^{\prime })\nabla _{d}^{\prime }\mathbf{%
\vekt{u}%
}_{b}(%
\vekt{x}%
^{\prime })  \label{2.17}
\end{equation}%
\begin{equation}
S_{abcd}(%
\vekt{x}%
,%
\vekt{x}%
^{\prime })=\beta \int_{0}^{\infty }dt\langle \lbrack \func{e}^{(1-\mathcal{P%
})i\mathcal{L}t}\widehat{s}_{ac}(%
\vekt{x}%
)]\widehat{s}_{bd}(%
\vekt{x}%
^{\prime }\rangle _{L}  \label{2.18}
\end{equation}

It is seen immediately that (\ref{2.16}) is the hydrodynamic velocity
equation for stationary incompressible flow. The stress Tensor $%
\vekt{R}%
$ is, in general, a nonlinear functional of $\mathbf{%
\vekt{u}%
}$. Grabert, in \cite{gr}, sec. 8.5, performs an approximation to the first
order of $\ \mathbf{\nabla 
\vekt{u}%
}$ and obtains exactly the Stokes form of the tensor. Therefore, in this
approximation, (\ref{2.16}) is the stationary Navier-Stokes equation.

By (\ref{2.14}), $\mathbf{%
\vekt{u}%
}$ is essentially equal to the expectation of the momentum density. By
definition from probability theory, the expectation ist an average built
from a set of realisations of the process, which can be interpreted as
repetitions of the experiment und identical external conditions. If the flow
exhibits macroscopic, i. e. turbulent, fluctuations, the averaging process
includes these. There is barely another possibility than, in this case, to
interpret $\mathbf{%
\vekt{u}%
}$\ as the \textit{mean} velocity of the flow. On the other hand, if one
wants to define the 'point' velocity of turbulent flow, it would be
necessary to introduce a conditional expectation, which excludes the
macroscopic fluctuations, with respect of which the mean quantity would
still be a random variable. - The preceding statement is quite general; it
is still valid if one performs a projection operator analyses. Thus, in case
of stationary turbulent flow (\ref{2.16}) is the \textit{mean velocity}
equation. It is of course necessary to check this theoretical statement by
bringing the equation into a form that can be evaluated, and comparing the
results with a turbulent flow experiment.

The difference is that for turbulent flow the nonlinear part of the stress
tensor $%
\vekt{R}%
$ is essential. The definition formula for the stress tensor kernel function 
$%
\vekt{S}%
$ (\ref{2.18}) contains a time correlation function in local equilibrium.
This is a quantity which, for processes with constant mass density and
temperature, is a functional of the velocity field. It has to be evaluated
in advance, by a separate statistical-mechanical formalism. At present there
is no theoretical means to perform this; instead, it is possible to
calculate correlation functions for \textit{total} equilibrium. It should be
emphasized that, since total equilibrium corresponds to macroscopic rest,
the latter quantities do not depend on the flow properties; they are
material 'constants' of the fluid. - In order to pursue the analysis, $%
\vekt{S}%
$ has been expanded into a functional power series in $%
\vekt{u}%
$; as will be seen, the coefficients of the series contain total equilibrium
correlations. It will be necessary then to evaluate the lowest order terms
of the expansion and to work with formulas \ (\ref{2.16}), (\ref{2.17})\
with an approximated quantity $%
\vekt{S}%
$. When the formlism is applied to a given flow configuration, it is
possible to render all variable quantities in the expansion dimension-free,
which causes certain constant factors to appear in the terms of it. The
author elaborates on an application to circular jet flow; where these
factors show up actually as increasing powers of the Reynolds number $Re$.
Thus, it is seen that in this case the application should be restricted to
low Reynolds number flow, just beyond the laminar-turbulent transition.

\section{\protect\bigskip Expansion of the kernel function}

\setcounter{equation}{0}

$%
\vekt{S}%
$ depends on $%
\vekt{u}%
$ via the quantities $%
\vekt{b}%
$ (\ref{2.6b}) in the formula for the local equilibrium distribution. The
series expansion of $%
\vekt{S}%
$\ is performed by expanding it with respect to the $%
\vekt{b}%
$\ at the point $%
\vekt{b}%
=%
\vekt{b}%
_{0}$ which corresponds to $\mathbf{%
\vekt{u}%
=}0$: 
\begin{equation}
\vekt{b}%
_{0}=\{-\beta \mu \,,\,\beta \,,\,0\}  \label{3.1}
\end{equation}

\begin{equation}
\vekt{b}%
-%
\vekt{b}%
_{0}=\{\beta \frac{m}{2}u^{2}\,,\,0\,,\,-\beta \mathbf{%
\vekt{u}%
\,\}}  \label{3.2}
\end{equation}

Replacing $%
\vekt{b}%
$\ by $%
\vekt{b}%
_{0}$\ changes the local equilibrium distribution (\ref{2.6a}) into the
total equilibrium distribution (\ref{2.5a}) with the prescribed mass density
and temperature. The power expansion of $%
\vekt{S}%
$ reads:

\begin{multline}
\vekt{S}%
=%
\vekt{S}%
|_{%
\vekt{b}%
_{0}}+\frac{\delta 
\vekt{S}%
}{\delta 
\vekt{b}%
}|_{%
\vekt{b}%
_{0}}\ast (%
\vekt{b}%
-%
\vekt{b}%
_{0})+\frac{1}{2!}\frac{\delta ^{2}%
\vekt{S}%
}{\delta 
\vekt{b}%
\,\delta 
\vekt{b}%
}|_{%
\vekt{b}%
_{0}}\ast \ast \{(%
\vekt{b}%
-%
\vekt{b}%
_{0}),(%
\vekt{b}%
-%
\vekt{b}%
_{0})\}+\cdots \\
=%
\vekt{S}%
^{(0)}+%
\vekt{S}%
^{(1)}+%
\vekt{S}%
^{(2)}+\cdots  \label{3.3}
\end{multline}

In the second row, symbols are applied to the different orders of the
expansion. The expansion in $%
\vekt{u}%
$\ is obtained from (\ref{3.3}) by restricting the summation inherent in the 
$\ast $-operation to the last element of $%
\vekt{b}%
-%
\vekt{b}%
_{0}$ in (\ref{3.2}). It is seen there that the first element also depends
on $%
\vekt{u}%
$. But for the present investigation, where no sound or heat conduction
processes are considered, the expansion coefficients come out to be non-zero
only for 'Latin' values of the indices; thus the first element of $%
\vekt{b}%
$\ does not contribute.

The linear part of $%
\vekt{R}%
$\ which results from the constant term $%
\vekt{S}%
^{(0)}$\ of the expansion (\ref{3.3}) leads to the Stokes form of the stress
tensor, which coincides with the result of Grabert. At present, the 2$^{%
\text{nd}}$ and 3$^{\text{rd}}$ order terms of $%
\vekt{R}%
$ have been analyzed. From (\ref{2.17}) it can be seen that these\ stem from
the parts $%
\vekt{S}%
^{(1)}$ and $%
\vekt{S}%
^{(2)}$ of the $%
\vekt{S}%
$-series, respectively. To obtain these formulas, it is necessary to
calculate the first and second order functional derivatives of $%
\vekt{S}%
$. The first of these has been done in Piest \cite{pi89}. For completeness,
the derivation is repeated in the appendix. The result is formula (\ref{aa17}%
); by setting $%
\vekt{b}%
=%
\vekt{b}%
_{0}$ one obtains:%
\begin{multline}
\frac{\delta S_{abcd}(%
\vekt{x}%
,%
\vekt{x}%
^{\prime })}{\delta b_{e}(%
\vekt{x}%
^{\prime \prime })\,}|_{%
\vekt{b}%
=%
\vekt{b}%
_{0}} \\
=-\beta \int_{0}^{\infty }dt\langle \lbrack \func{e}^{(1-\mathcal{P}_{0})i%
\mathcal{L}t}(1-\mathcal{P}_{0})s_{ac}(%
\vekt{x}%
)][(1-\mathcal{P}_{0})s_{bd}(%
\vekt{x}%
^{\prime })]p_{e}(%
\vekt{x}%
^{\prime \prime })\rangle _{0}  \label{3.4}
\end{multline}

$\mathcal{P}_{0}$ is the total equilibrium projection operator:%
\begin{equation}
\mathcal{P}_{0}g=\langle g\rangle _{0}+\langle g\,\delta _{0}%
\vekt{a}%
\rangle _{0}\ast \langle \delta _{0}%
\vekt{a}%
\,\delta _{0}%
\vekt{a}%
\rangle _{0}^{-1}\ast \delta _{0}%
\vekt{a}
\label{3.5}
\end{equation}

Here, $\delta _{0}g=g-\left\langle g\right\rangle _{0}$, and we have $%
\left\langle 
\vekt{p}%
\right\rangle _{0}=0$.- The total equilibrium triple correlation function
contained in (\ref{3.4}) can be calculated using the mode coupling technique
in the form of Martin et al. \cite{msr73}, which is a systematic version of
the method of Kawasaki \cite{k61}, and has been complemented by Deker and
Haake \cite{dh75}. The formulas have been evaluated by the author, and
applied using the experimental data of the circular jet (Piest \cite{pi90}).
One obtains the solution again in form of a power series. The expansion
parameter $\chi $ reads: 
\begin{equation}
\chi =\frac{1}{\beta \nu ^{2}\rho x}  \label{3.6}
\end{equation}

$\nu $ is the kinematic viscosity, $x$ the distance of the observation point
from the orifice. With a typical value $x=0.3\,\left[ \text{m}\right] $, for
a laboratory experiment in air, one obtains $\chi =5.8\cdot 10^{-11}$, which
is so small that practically only the zero order term of the expansion
counts.- It was rather surprising that for the linear term of the $\mathbf{%
\vekt{u}%
}$-series of $%
\vekt{S}%
$ (\ref{3.3}), a zeroth order $\chi $-Term is found which leads to a second
order term of the friction force $%
\vekt{D}%
=\mathbf{\nabla }\cdot 
\vekt{R}%
$: 
\begin{equation}
\vekt{D}%
^{(2,0)}=\rho (\mathbf{%
\vekt{u}%
\cdot \nabla 
\vekt{u}%
)-}\frac{\lambda }{2}\mathbf{\nabla (%
\vekt{u}%
}\cdot \mathbf{%
\vekt{u}%
)}  \label{3.7}
\end{equation}

$\lambda =\alpha /(\gamma c_{V})$ is a physical parameter of the fluid, $%
\alpha $ being the thermal expansion coefficient, $\gamma $ the isothermal
compressibility, and $c_{V}$ the specific heat for constant volume. The
apparent existence of this term poses a problem to the intepretation of
equation (\ref{2.16}). It should be noticed that, since for sufficient low
Reynolds number the flow is laminar, there are no macroscopic fluctuations
then, and $\mathbf{%
\vekt{u}%
}$\ equals the classical hydrodynamic 'point' velocity. Therefore, for
decreasing $Re$, equation (\ref{2.16}) should reduce to the stationary
Navier-Stokes equation. On the other hand, $%
\vekt{D}%
^{(2,0)}$ is of equal power in $Re$\ as the quadratic term on the left of (%
\ref{2.16}) so that, even for small $Re$, (\ref{2.16})\ remains different.
Since this is not possible, something must be wrong with the prerequisites
of the derivation of (\ref{3.7}). As a preliminary measure, it is assumed
that the term actually does not exist. It is of course one of the most
urging requirements for this approach to sufficiently explain this defect.

The third order term of the expansion of $%
\vekt{R}%
$ contains, after (\ref{2.17}) and (\ref{3.3}), the second-order derivative,
which is presented in (\ref{aa35}). When this formula is applied to $%
\vekt{b}%
=%
\vekt{b}%
_{0}$ as in (\ref{3.4}), the summation in the corresponding term in (\ref%
{3.3}) is restricted to the last term of (\ref{3.2}) and the result is
inserted into (\ref{2.17}), on obtains for the friction force $%
\vekt{D}%
=\mathbf{\nabla }\cdot 
\vekt{R}%
$:

\begin{equation}
D_{a}^{(3)}(%
\vekt{x}%
)=\int d%
\vekt{x}%
^{\prime }d%
\vekt{x}%
^{\prime \prime }d%
\vekt{x}%
^{\prime \prime \prime }K_{abcd}(%
\vekt{x}%
,%
\vekt{x}%
^{\prime },%
\vekt{x}%
^{\prime \prime },%
\vekt{x}%
^{\prime \prime \prime })\mathbf{%
\vekt{u}%
}_{b}(%
\vekt{x}%
^{\prime })\mathbf{%
\vekt{u}%
}_{c}(%
\vekt{x}%
^{\prime \prime })\mathbf{%
\vekt{u}%
}_{d}(%
\vekt{x}%
^{\prime \prime \prime })  \label{3.8}
\end{equation}%
\begin{multline}
K_{abcd}(%
\vekt{x}%
,%
\vekt{x}%
^{\prime },%
\vekt{x}%
^{\prime \prime },%
\vekt{x}%
^{\prime \prime \prime })=-\frac{1}{2}\beta ^{3}\int_{0}^{\infty }dt\langle
\lbrack \func{e}^{(1-\mathcal{P}_{0})i\mathcal{L}t}\nabla _{e}(1-\mathcal{P}%
_{0})s_{ae}(%
\vekt{x}%
)]\times \\
\times \lbrack \nabla _{f}^{\prime }(1-\mathcal{P}_{0})s_{bf}(%
\vekt{x}%
^{\prime })]\delta _{0}(p_{c}(%
\vekt{x}%
^{\prime \prime })p_{d}(%
\vekt{x}%
^{\prime \prime \prime }))\rangle _{0}  \label{3.9}
\end{multline}

For formal reasons, all $\nabla $-operations have been transferred to the
correlation function, partly by partial integration. Formulas (\ref{3.8}), (%
\ref{3.9}) are the main result of the present paper. The definition formula
for the kernel function $%
\vekt{K}%
$ contains a time integral over a correlation function which is double in
time but quadruple in space. To calculate it, the author has again applied
the technique described in Martin et al. \cite{msr73}, Deker and Haake \cite%
{dh75}; it had to be enlarged slightly so that quadruple correlations could
be determined. This investigation will be presented in a separate paper. The
resulting formula is rather lenghty and will not be given here. It should be
emphasized that with $%
\vekt{K}%
$\ obtained as an explicit formula of the four space variables, (\ref{2.16}%
), (\ref{3.8})\ form a closed system for calculating the velocity field for
a given flow configuration. Moreover, since this formula is the \textit{%
lowest} order term of the expansion (see the comment to (\ref{3.7})), (\ref%
{2.16}), (\ref{3.8}) is the \textit{simplest} form of the system for
checking, by comparing with experimental results, whether the approach
works. As has been mentioned, a numerical test using the data for the
circular jet is in progress.

\section{Summary}

An approach to arrive at the mean velocity equation for turbulent fluid flow
has been attempted with the aid of the projection operator technique in
classical Statistical Mechanics. The hydrodynamic velocity is defined in
this technique via the conjugate thermodynamic fields in the formula for the
relevant probability density; multiplied by the mass density, it is
identical to the expectation of the microscopic momentum density. It is
argued that in situations where the fluid flow is turbulent, this is
precisely the mean velocity field of the flow. If this argument is correct,
the momentum component of the generalized transport equation derived by this
technique is the mean velocity equation.

Stationary incompressible flow for constant mass density and temperature is
considered. The formula for the stress tensor is a nonlinear functional of
the velocity, the linear part of which has the form of the Stokes tensor.
The formula containes a local equilibrium time correlation function. At
present, there exists a theory for calculating correlation functions for
total equilibrium only. As a preliminary measure, the stress tensor has been
developed into a power series in the velocity, though this limits the
applicability of the equation to low Reynolds number flow. The coefficients
of the expansion contain total equilibrium correlations which can be
calculated. The second order term has been evaluated in a former paper of
the author. For completeness, the main results have been reported here. The
calculation leads to a second order friction term which is comparable in
form to the convolution term of the equation. This constitutes a problem to
the present approach since from general knowledge about the Reynolds
equation such a term cannot appear.

For the third order term, the form of the kernel function has been
calculated. The formula has been evaluated with the aid of the mode coupling
theory of Statistical Mechanics; the results will be reported in a separate
paper. A numerical calculation in order to test the equation with the
experimental data for the circular jet is under way.

\appendix%

\section*{Appendix:\protect\linebreak Calculation of functional derivatives}

\setcounter{equation}{0}

\renewcommand{\theequation}{a.\arabic{equation}}%

In this appendix, the first two derivatives, with respect to $%
\vekt{b}%
$, of the kernel function $%
\vekt{S}%
$ (\ref{2.18}) are calculated. The right hand side of (\ref{2.18}) depends
on $%
\vekt{b}%
$ at four different places; thus we may write:%
\begin{equation}
\frac{\delta S_{abcd}(%
\vekt{x}%
,%
\vekt{x}%
^{\prime })}{\delta b_{e}(%
\vekt{x}%
^{\prime \prime })}=\sum_{i=1}^{4}\frac{\delta 
\vekt{S}%
}{\delta 
\vekt{b}%
\,}^{(i)}  \label{aa1}
\end{equation}

The definition of the four terms is given in the formulas to follow. In
addition, certain auxiliary formulas are written which can be verified
directly. For the first term, we use the rule:%
\begin{equation}
\frac{\delta f_{L}}{\delta 
\vekt{b}%
}=-f_{L}\delta 
\vekt{a}
\label{aa2}
\end{equation}

The first term reads: 
\begin{multline}
\frac{\delta 
\vekt{S}%
}{\delta 
\vekt{b}%
\,}^{(1)}=\beta \int_{0}^{\infty }dt\,tr\{\frac{\delta f_{L}}{\delta b_{f}(%
\vekt{x}%
^{\prime \prime \prime })}[\func{e}^{(1-\mathcal{P})i\mathcal{L}t}\widehat{s}%
_{ac}(%
\vekt{x}%
)]\widehat{s}_{bd}(%
\vekt{x}%
^{\prime })\} \\
=-\beta \int_{0}^{\infty }dt\langle \lbrack \func{e}^{(1-\mathcal{P})i%
\mathcal{L}t}\widehat{s}_{ac}(%
\vekt{x}%
)]\widehat{s}_{bd}(%
\vekt{x}%
^{\prime })\delta p_{e}(%
\vekt{x}%
^{\prime \prime })\rangle _{L}  \label{aa3}
\end{multline}

Next, we need the formula for the differentiation of the projection operator 
$\mathcal{P}$ : 
\begin{equation}
\frac{\delta \mathcal{P}}{\delta 
\vekt{b}%
}=-\mathcal{P\,}\delta 
\vekt{a}%
(1-\mathcal{P)}  \label{aa4}
\end{equation}

The second term reads:

\begin{multline}
\frac{\delta 
\vekt{S}%
}{\delta 
\vekt{b}%
\,}^{(2)}=-\beta \int_{0}^{\infty }dt\langle \lbrack \func{e}^{(1-\mathcal{P}%
)i\mathcal{L}t}\frac{\delta \mathcal{P}}{\delta 
\vekt{b}%
_{e}(%
\vekt{x}%
^{\prime \prime })}s_{ac}(%
\vekt{x}%
)]\widehat{s}_{bd}(%
\vekt{x}%
^{\prime })\rangle _{L} \\
=\beta \int_{0}^{\infty }dt\langle \lbrack \func{e}^{(1-\mathcal{P})i%
\mathcal{L}t}\mathcal{P}\widehat{s}_{ac}(%
\vekt{x}%
)\delta p_{e}(%
\vekt{x}%
^{\prime \prime })]\widehat{s}_{bd}(%
\vekt{x}%
^{\prime })\rangle _{L}  \label{aa5}
\end{multline}

When (\ref{aa4}) is applied to the third term, one finds, after some
manipulations: 
\begin{multline}
\frac{\delta 
\vekt{S}%
}{\delta 
\vekt{b}%
\,}^{(3)}=-\beta \int_{0}^{\infty }dt\langle \lbrack \func{e}^{(1-\mathcal{P}%
)i\mathcal{L}t}\widehat{s}_{ac}(%
\vekt{x}%
)]\frac{\delta \mathcal{P}}{\delta 
\vekt{b}%
_{e}(%
\vekt{x}%
^{\prime \prime })}s_{bd}(%
\vekt{x}%
^{\prime })\rangle _{L} \\
=\beta \int_{0}^{\infty }dt\langle \lbrack (1-\mathcal{P})\func{e}^{(1-%
\mathcal{P})i\mathcal{L}t}\widehat{s}_{ac}(%
\vekt{x}%
)]\mathcal{P}\widehat{s}_{bd}(%
\vekt{x}%
^{\prime })\delta p_{e}(%
\vekt{x}%
^{\prime \prime })\rangle _{L} \\
=0  \label{aa6}
\end{multline}

In the second step, it is used that we have:%
\begin{equation}
\func{e}^{(1-\mathcal{P})i\mathcal{L}t}(1-\mathcal{P})=(1-\mathcal{P})\func{e%
}^{(1-\mathcal{P})i\mathcal{L}t}(1-\mathcal{P})  \label{aa7}
\end{equation}%
\begin{equation}
\langle \lbrack (1-\mathcal{P})g_{1}]\mathcal{P}g_{2}\rangle _{L}=0
\label{aa8}
\end{equation}

(\ref{aa8}) is valid for any microscopic functions $g_{1},\,g_{2}$ . - The
part containing the differentiation of the exponential operator is a
somewhat more involved. The differentiation formula is: 
\begin{equation}
\frac{\delta \func{e}^{(1-\mathcal{P})i\mathcal{L}t}}{\delta 
\vekt{b}%
}=\int_{0}^{t}dt^{\prime }\func{e}^{(1-\mathcal{P})i\mathcal{L}t^{\prime }}%
\mathcal{P\,\delta }%
\vekt{a}%
\frac{d}{dt}\func{e}^{(1-\mathcal{P})i\mathcal{L(}t-t^{\prime })}
\label{aa9}
\end{equation}

For the corresponding time integral, it follows after interchanging the
succession of integrations:%
\begin{multline}
\int_{0}^{\infty }dt\frac{\delta \func{e}^{(1-\mathcal{P})i\mathcal{L}t}}{%
\delta 
\vekt{b}%
_{f}(%
\vekt{x}%
^{\prime \prime \prime })}=\int_{0}^{\infty }dt^{\prime }\int_{0}^{\infty }dt%
\func{e}^{(1-\mathcal{P})i\mathcal{L}t^{\prime }}\mathcal{P\,}\delta p_{e}(%
\vekt{x}%
^{\prime \prime })\frac{d}{dt}\func{e}^{(1-\mathcal{P})i\mathcal{L}t} \\
=\int_{0}^{\infty }dt^{\prime }\{\lim_{t\rightarrow \infty }\func{e}^{(1-%
\mathcal{P})i\mathcal{L}t^{\prime }}\mathcal{P\,}\delta p_{e}(%
\vekt{x}%
^{\prime \prime })\func{e}^{(1-\mathcal{P})i\mathcal{L}t}-\func{e}^{(1-%
\mathcal{P})i\mathcal{L}t^{\prime }}\mathcal{P\,}\delta p_{e}(%
\vekt{x}%
^{\prime \prime })\}  \label{aa10}
\end{multline}

One obtains for the fourth term: 
\begin{multline}
\frac{\delta 
\vekt{S}%
}{\delta 
\vekt{b}%
\,}^{(4)}=\beta \int_{0}^{\infty }dt\langle \lbrack \frac{\delta \func{e}%
^{(1-\mathcal{P})i\mathcal{L}t}}{\delta 
\vekt{b}%
_{e}(%
\vekt{x}%
^{\prime \prime })}\widehat{s}_{ac}(%
\vekt{x}%
)]\widehat{s}_{bd}(%
\vekt{x}%
^{\prime })\rangle _{L} \\
=\beta \int_{0}^{\infty }dt^{\prime }\{\lim_{t\rightarrow \infty }\langle
\lbrack \func{e}^{(1-\mathcal{P})i\mathcal{L}t^{\prime }}\mathcal{P\,}\delta
p_{e}(%
\vekt{x}%
^{\prime \prime })\func{e}^{(1-\mathcal{P})i\mathcal{L}t}\widehat{s}_{ac}(%
\vekt{x}%
)]\widehat{s}_{bd}(%
\vekt{x}%
^{\prime })\rangle _{L} \\
-\langle \lbrack \func{e}^{(1-\mathcal{P})i\mathcal{L}t^{\prime }}\mathcal{%
P\,}\delta p_{e}(%
\vekt{x}%
^{\prime \prime })\widehat{s}_{ac}(%
\vekt{x}%
)]\widehat{s}_{bd}(%
\vekt{x}%
^{\prime })\rangle _{L}\}  \label{aa11}
\end{multline}

We have to evaluate the limit expression. We will \textit{assume} here that
in the limit of large times, the factors of a time correlation will become
statistically independent so that for stationary processes we have: 
\begin{equation}
\lim_{t\rightarrow \infty }\langle \lbrack \func{e}^{(1-\mathcal{P})i%
\mathcal{L}t}A]B\rangle _{L}=\langle A\rangle _{L}\langle B\rangle _{L}
\label{aa12}
\end{equation}

For any phase space function $g$, let us define a quantity $F(t^{\prime })$:%
\begin{equation}
F(t^{\prime })=\lim_{t\rightarrow \infty }\langle \lbrack \func{e}^{(1-%
\mathcal{P})i\mathcal{L}t^{\prime }}\mathcal{P\,}\delta p_{e}(%
\vekt{x}%
^{\prime \prime })\func{e}^{(1-\mathcal{P})i\mathcal{L}t}\widehat{s}_{ac}(%
\vekt{x}%
)]g\rangle _{L}  \label{aa13}
\end{equation}

The evaluation results in: 
\begin{multline}
F(t^{\prime })=\lim_{t\rightarrow \infty }\{\langle \mathcal{\,}\delta p_{e}(%
\vekt{x}%
^{\prime \prime })\func{e}^{(1-\mathcal{P})i\mathcal{L}t}\widehat{s}_{ac}(%
\vekt{x}%
)\rangle _{L}\langle g\rangle _{L} \\
+\langle \mathcal{\,}\delta p_{e}(%
\vekt{x}%
^{\prime \prime })[\func{e}^{(1-\mathcal{P})i\mathcal{L}t}\widehat{s}_{ac}(%
\vekt{x}%
)]\delta 
\vekt{a}%
\rangle _{L}\ast \langle \delta 
\vekt{a}%
\,\delta 
\vekt{a}%
\rangle _{L}^{-1}\ast \langle \lbrack \func{e}^{(1-\mathcal{P})i\mathcal{L}%
t^{\prime }}\delta 
\vekt{a}%
]g\rangle _{L}\} \\
=\langle \widehat{s}_{ac}(%
\vekt{x}%
)\rangle _{L}\{\langle \mathcal{\,}\delta p_{e}(%
\vekt{x}%
^{\prime \prime })\rangle _{L}\langle g\rangle _{L}+\langle \mathcal{\,}%
\delta p_{e}(%
\vekt{x}%
^{\prime \prime })\delta 
\vekt{a}%
\rangle _{L}\ast \\
\ast \langle \delta 
\vekt{a}%
\,\delta 
\vekt{a}%
\rangle _{L}^{-1}\ast \langle \lbrack \func{e}^{(1-\mathcal{P})i\mathcal{L}%
t^{\prime }}\delta 
\vekt{a}%
]g\rangle _{L}\}=0  \label{aa14}
\end{multline}

The first step is the evaluation of $\mathcal{P}$ using (\ref{2.8}); next,
we use (\ref{aa12}); finally, the factor outside the curled brackets
vanishes, since we have, for any phase space function $g$:%
\begin{equation}
\langle (1-\mathcal{P})g\rangle _{L}=0  \label{aa15}
\end{equation}

Therefore, in (\ref{aa11}), the limit term vanishes, and we have:\ 
\begin{equation}
\frac{\delta 
\vekt{S}%
}{\delta 
\vekt{b}%
\,}^{(4)}=\beta \int_{0}^{\infty }dt\langle \lbrack \func{e}^{(1-\mathcal{P}%
)i\mathcal{L}t}\mathcal{P}\widehat{s}_{ac}(%
\vekt{x}%
)\delta p_{e}(%
\vekt{x}%
^{\prime \prime })]\widehat{s}_{bd}(%
\vekt{x}%
^{\prime })\delta p_{e}(%
\vekt{x}%
^{\prime \prime })\rangle _{L}=-\frac{\delta 
\vekt{S}%
}{\delta 
\vekt{b}%
\,}^{(2)}  \label{aa16}
\end{equation}

Thus, we obtain: 
\begin{equation}
\frac{\delta S_{abcd}(%
\vekt{x}%
,%
\vekt{x}%
^{\prime })}{\delta b_{e}(%
\vekt{x}%
^{\prime \prime })\,}=\frac{\delta 
\vekt{S}%
}{\delta 
\vekt{b}%
\,}^{(1)}=-\beta \int_{0}^{\infty }dt\langle \lbrack \func{e}^{(1-\mathcal{P}%
)i\mathcal{L}t}\widehat{s}_{ac}(%
\vekt{x}%
)]\widehat{s}_{bd}(%
\vekt{x}%
^{\prime })\delta p_{e}(%
\vekt{x}%
^{\prime \prime })\rangle _{L}  \label{aa17}
\end{equation}

The calculation of the coefficients of the second order derivative of $%
\vekt{S}%
$ parallels to a certain extent that of the first order. The right hand side
of (\ref{aa17}) depends on $%
\vekt{b}%
$ at five different places; thus we write:%
\begin{equation}
\frac{\delta ^{2}S_{abcd}(%
\vekt{x}%
,%
\vekt{x}%
^{\prime })}{\delta b_{e}(%
\vekt{x}%
^{\prime \prime })\,\delta b_{f}(%
\vekt{x}%
^{\prime \prime \prime })}=\sum_{i=1}^{5}\frac{\delta ^{2}%
\vekt{S}%
}{\delta 
\vekt{b}%
\,\delta 
\vekt{b}%
}^{(i)}  \label{aa18}
\end{equation}

\bigskip For the first part, we use (\ref{aa2}):%
\begin{multline}
\frac{\delta ^{2}%
\vekt{S}%
}{\delta 
\vekt{b}%
\,\delta 
\vekt{b}%
}^{(1)}=-\beta \int_{0}^{\infty }dt\,tr\{\frac{\delta f_{L}}{\delta b_{f}(%
\vekt{x}%
^{\prime \prime \prime })}[\func{e}^{(1-\mathcal{P})i\mathcal{L}t}\widehat{s}%
_{ac}(%
\vekt{x}%
)]\widehat{s}_{bd}(%
\vekt{x}%
^{\prime })\delta p_{e}(%
\vekt{x}%
^{\prime \prime })\} \\
=\beta \int_{0}^{\infty }dt\langle \lbrack \func{e}^{(1-\mathcal{P})i%
\mathcal{L}t}\widehat{s}_{ac}(%
\vekt{x}%
)]\widehat{s}_{bd}(%
\vekt{x}%
^{\prime })\delta p_{e}(%
\vekt{x}%
^{\prime \prime })\delta p_{f}(%
\vekt{x}%
^{\prime \prime \prime })\rangle _{L}  \label{aa19}
\end{multline}

The second term , with (\ref{aa4}), turns out to be:

\begin{multline}
\frac{\delta ^{2}%
\vekt{S}%
}{\delta 
\vekt{b}%
\,\delta 
\vekt{b}%
}^{(2)}=\beta \int_{0}^{\infty }dt\langle \lbrack \func{e}^{(1-\mathcal{P})i%
\mathcal{L}t}\frac{\delta \mathcal{P}}{\delta 
\vekt{b}%
_{f}(%
\vekt{x}%
^{\prime \prime \prime })}s_{ac}(%
\vekt{x}%
)]\widehat{s}_{bd}(%
\vekt{x}%
^{\prime })\delta p_{e}(%
\vekt{x}%
^{\prime \prime })\rangle _{L} \\
=-\beta \int_{0}^{\infty }dt\langle \lbrack \func{e}^{(1-\mathcal{P})i%
\mathcal{L}t}\mathcal{P}\widehat{s}_{ac}(%
\vekt{x}%
)\delta p_{f}(%
\vekt{x}%
^{\prime \prime \prime })]\widehat{s}_{bd}(%
\vekt{x}%
^{\prime })\delta p_{e}(%
\vekt{x}%
^{\prime \prime })\rangle _{L}  \label{aa20}
\end{multline}

In the same way, we have for the third term:

\begin{multline}
\frac{\delta ^{2}%
\vekt{S}%
}{\delta 
\vekt{b}%
\,\delta 
\vekt{b}%
}^{(3)}=\beta \int_{0}^{\infty }dt\langle \lbrack \func{e}^{(1-\mathcal{P})i%
\mathcal{L}t}\widehat{s}_{ac}(%
\vekt{x}%
)]\frac{\delta \mathcal{P}}{\delta 
\vekt{b}%
_{f}(%
\vekt{x}%
^{\prime \prime \prime })}s_{bd}(%
\vekt{x}%
^{\prime })\delta p_{e}(%
\vekt{x}%
^{\prime \prime })\rangle _{L} \\
=-\beta \int_{0}^{\infty }dt\langle \lbrack \func{e}^{(1-\mathcal{P})i%
\mathcal{L}t}\widehat{s}_{ac}(%
\vekt{x}%
)][\mathcal{P}\widehat{s}_{bd}(%
\vekt{x}%
^{\prime })\delta p_{f}(%
\vekt{x}%
^{\prime \prime \prime })]\delta p_{e}(%
\vekt{x}%
^{\prime \prime })\rangle _{L}  \label{aa21}
\end{multline}

\bigskip For the fourth part containing the differentiation of the
exponential operator we have, with (\ref{aa10}): 
\begin{multline}
\frac{\delta ^{2}%
\vekt{S}%
}{\delta 
\vekt{b}%
\,\delta 
\vekt{b}%
}^{(4)}=-\beta \int_{0}^{\infty }dt\langle \lbrack \frac{\delta \func{e}^{(1-%
\mathcal{P})i\mathcal{L}t}}{\delta 
\vekt{b}%
_{f}(%
\vekt{x}%
^{\prime \prime \prime })}\widehat{s}_{ac}(%
\vekt{x}%
)]\widehat{s}_{bd}(%
\vekt{x}%
^{\prime })\delta p_{e}(%
\vekt{x}%
^{\prime \prime })\rangle _{L} \\
=-\beta \int_{0}^{\infty }dt^{\prime }\{\lim_{t\rightarrow \infty }\langle
\lbrack \func{e}^{(1-\mathcal{P})i\mathcal{L}t^{\prime }}\mathcal{P\,}\delta
p_{f}(%
\vekt{x}%
^{\prime \prime \prime })\func{e}^{(1-\mathcal{P})i\mathcal{L}t}\widehat{s}%
_{ac}(%
\vekt{x}%
)]\widehat{s}_{bd}(%
\vekt{x}%
^{\prime })\delta p_{e}(%
\vekt{x}%
^{\prime \prime })\rangle _{L} \\
-\langle \lbrack \func{e}^{(1-\mathcal{P})i\mathcal{L}t^{\prime }}\mathcal{%
P\,}\delta p_{f}(%
\vekt{x}%
^{\prime \prime \prime })\widehat{s}_{ac}(%
\vekt{x}%
)]\widehat{s}_{bd}(%
\vekt{x}%
^{\prime })\delta p_{e}(%
\vekt{x}%
^{\prime \prime })\rangle _{L}\}  \label{aa22}
\end{multline}

The evaluation parallels that of the corresponding term of the first
derivative. We finally find: \ 
\begin{equation}
\frac{\delta ^{2}%
\vekt{S}%
}{\delta 
\vekt{b}%
\,\delta 
\vekt{b}%
}^{(4)}=\beta \int_{0}^{\infty }dt\langle \lbrack \func{e}^{(1-\mathcal{P})i%
\mathcal{L}t}\mathcal{P}\widehat{s}_{ac}(%
\vekt{x}%
)\delta p_{f}(%
\vekt{x}%
^{\prime \prime \prime })]\widehat{s}_{bd}(%
\vekt{x}%
^{\prime })\delta p_{e}(%
\vekt{x}%
^{\prime \prime })\rangle _{L}=-\frac{\delta ^{2}%
\vekt{S}%
}{\delta 
\vekt{b}%
\,\delta 
\vekt{b}%
}^{(2)}  \label{aa23}
\end{equation}

For the fifth term, we need the differentiation rule for $\langle 
\vekt{a}%
$ $\rangle _{L}$:%
\begin{equation}
\frac{\delta \langle 
\vekt{a}%
\rangle _{L}}{\delta 
\vekt{b}%
}=-\langle \delta 
\vekt{a}%
\;\delta 
\vekt{a}%
\rangle _{L}  \label{aa24}
\end{equation}

We obtain: 
\begin{multline}
\frac{\delta ^{2}%
\vekt{S}%
}{\delta 
\vekt{b}%
\,\delta 
\vekt{b}%
}^{(5)}=\beta \int_{0}^{\infty }dt\langle \lbrack \func{e}^{(1-\mathcal{P})i%
\mathcal{L}t}\widehat{s}_{ac}(%
\vekt{x}%
)]\widehat{s}_{bd}(%
\vekt{x}%
^{\prime })\rangle _{L}\frac{\delta \langle p_{e}(%
\vekt{x}%
^{\prime \prime }\rangle _{L}}{\delta 
\vekt{b}%
_{f}(%
\vekt{x}%
^{\prime \prime \prime })} \\
=-\beta \int_{0}^{\infty }dt\langle \lbrack \func{e}^{(1-\mathcal{P})i%
\mathcal{L}t}\widehat{s}_{ac}(%
\vekt{x}%
)]\widehat{s}_{bd}(%
\vekt{x}%
^{\prime })\rangle _{L}\langle \delta p_{e}(%
\vekt{x}%
^{\prime \prime })\mathcal{\,}\delta p_{f}(%
\vekt{x}%
^{\prime \prime \prime })\rangle _{L}  \label{aa25}
\end{multline}

Finally, we want to show that the term (\ref{aa21}) vanishes. The first step
is to evaluate $\mathcal{P}$ with the aid of (\ref{2.8}):

\begin{multline}
\frac{\delta ^{2}%
\vekt{S}%
}{\delta 
\vekt{b}%
\,\delta 
\vekt{b}%
}^{(3)}=-\beta \int_{0}^{\infty }dt\{\langle \widehat{s}_{bd}(%
\vekt{x}%
^{\prime })\delta p_{f}(%
\vekt{x}%
^{\prime \prime \prime })\rangle _{L}\langle \lbrack \func{e}^{(1-\mathcal{P}%
)i\mathcal{L}t}\widehat{s}_{ac}(%
\vekt{x}%
)]\delta p_{e}(%
\vekt{x}%
^{\prime \prime })\rangle _{L} \\
+\langle \widehat{s}_{bd}(%
\vekt{x}%
^{\prime })\delta p_{f}(%
\vekt{x}%
^{\prime \prime \prime })\delta 
\vekt{a}%
\rangle _{L}\ast \langle \delta 
\vekt{a}%
\,\delta 
\vekt{a}%
\rangle _{L}^{-1}\ast \langle \lbrack \func{e}^{(1-\mathcal{P})i\mathcal{L}t}%
\widehat{s}_{ac}(%
\vekt{x}%
)]\delta 
\vekt{a}%
\delta p_{e}(%
\vekt{x}%
^{\prime \prime })\rangle _{L}\}  \label{aa26}
\end{multline}

The first term on the right vanishes. Both factors are zero; e. g.: 
\begin{equation}
\langle \widehat{s}_{bd}(%
\vekt{x}%
^{\prime })\delta p_{f}(%
\vekt{x}%
^{\prime \prime \prime })\rangle _{L}=\langle s_{bd}(%
\vekt{x}%
^{\prime })(1-\mathcal{P})\delta p_{f}(%
\vekt{x}%
^{\prime \prime \prime })\rangle _{L}=0  \label{aa27}
\end{equation}

To investigate the second term of (\ref{aa26}), we introduce some auxiliary
functions:%
\begin{equation}
Z=\int_{0}^{\infty }dt\func{e}^{(1-\mathcal{P})i\mathcal{L}t}\widehat{s}_{ac}
\label{aa28}
\end{equation}

We have, with a suitable chosen $Y$:%
\begin{equation}
Z=(1-\mathcal{P})Y  \label{aa29}
\end{equation}

Therefore:%
\begin{equation}
\langle Z\rangle _{L}=0  \label{aa30}
\end{equation}

and:%
\begin{equation}
\langle Z\,\delta 
\vekt{a}%
\rangle _{L}=0  \label{aa31}
\end{equation}

Moreover, we write:%
\begin{equation}
\Xi =\langle Z\delta 
\vekt{a}%
\delta p_{e}\rangle _{L}  \label{aa32}
\end{equation}

We want to show $\Xi =0$. We consider the identity:%
\begin{equation}
\frac{\delta \langle Z\delta 
\vekt{a}%
\rangle _{L}}{\delta b_{e}}=\langle \frac{\delta Z}{\delta b_{e}}\delta 
\vekt{a}%
\rangle _{L}-\langle Z\rangle _{L}\frac{\delta \langle 
\vekt{a}%
\rangle _{L}}{\delta b_{e}}+\Xi  \label{aa33}
\end{equation}

The left hand side is zero because of (\ref{aa31}); so is the second term on
the right because of (\ref{aa30}). Moreover: 
\begin{multline}
\frac{\delta Z}{\delta b_{e}}=\int_{0}^{\infty }dt\{[\frac{\delta }{\delta
b_{e}}\func{e}^{(1-\mathcal{P})i\mathcal{L}t}]\widehat{s}_{ac}-\func{e}^{(1-%
\mathcal{P})i\mathcal{L}t}\frac{\delta \mathcal{P}}{\delta b_{e}}s_{ac}\} \\
=\int_{0}^{\infty }dt^{\prime }\lim_{t\rightarrow \infty }\func{e}^{(1-%
\mathcal{P})i\mathcal{L}t^{\prime }}\mathcal{P\,}\delta p_{e}\func{e}^{(1-%
\mathcal{P})i\mathcal{L}t}\widehat{s}_{ac}(%
\vekt{x}%
)  \label{aa34}
\end{multline}

The first step is by direct calculation. To the terms, we apply (\ref{aa9}),
(\ref{aa4}) respectively, to obtain the second step. Comparing this with (%
\ref{aa13}), we find that the first term on the right of (\ref{aa33}) is $%
F(t^{\prime })$ applied to $g=\delta 
\vekt{a}%
$, and is therefore zero. Thus, we have shown that actually (\ref{aa26})
vanishes. In total, we obtain from (\ref{aa18}), together with (\ref{aa19}),
(\ref{aa23}), (\ref{aa24}):

\begin{multline}
\frac{\delta ^{2}S_{abcd}(%
\vekt{x}%
,%
\vekt{x}%
^{\prime })}{\delta b_{e}(%
\vekt{x}%
^{\prime \prime })\,\delta b_{f}(%
\vekt{x}%
^{\prime \prime \prime })}=\frac{\delta ^{2}%
\vekt{S}%
}{\delta 
\vekt{b}%
\,\delta 
\vekt{b}%
}^{(1)}+\frac{\delta ^{2}%
\vekt{S}%
}{\delta 
\vekt{b}%
\,\delta 
\vekt{b}%
}^{(5)} \\
=\beta \int_{0}^{\infty }dt\langle \lbrack \func{e}^{(1-\mathcal{P})i%
\mathcal{L}t}\widehat{s}_{ac}(%
\vekt{x}%
)]\widehat{s}_{bd}(%
\vekt{x}%
^{\prime })\times \\
\times \{\delta p_{e}(%
\vekt{x}%
^{\prime \prime })\delta p_{f}(%
\vekt{x}%
^{\prime \prime \prime })-\langle \delta p_{e}(%
\vekt{x}%
^{\prime \prime })\delta p_{f}(%
\vekt{x}%
^{\prime \prime \prime })\rangle _{L}\}\rangle _{L}  \label{aa35}
\end{multline}

\end{document}